

\input mn
\input epsf

\let\sec=\section
\let\ssec=\subsection


\def\bigstrut{\vrule width0pt height0.35truecm}
\font\japit = cmti10 at 11truept

\def\ref{\parskip =0pt\par\noindent\hangindent\parindent
    \parskip =2ex plus .5ex minus .1ex}
\def\gs{\mathrel{\lower0.6ex\hbox{$\buildrel {\textstyle >}
 \over {\scriptstyle \sim}$}}}
\def\ls{\mathrel{\lower0.6ex\hbox{$\buildrel {\textstyle <}
 \over {\scriptstyle \sim}$}}}
\newcount\equationo
\equationo = 0
\def\leftdisplay#1$${\leftline{$\displaystyle{#1}$
  \global\advance\equationo by1\hfill (\the\equationo )}$$}
\everydisplay{\leftdisplay}

\def\hompc{\,h\;{\rm Mpc}^{-1}}
\def\mpcoh{\,h^{-1}\,{\rm Mpc}}

 

\def\apj{ApJ}
\def\apjs{ApJS}
\def\mn{MNRAS}
\def\nat{Nat}
\newcount\blogs
\newcount\fred
\newcount\bert
\newcount\berny
\newcount\ernie
\newcount\blah
\newcount\qdash
\newcount\sdash
\newcount\slong
\fred=0

\def\ineqn#1{\global #1 = \equationo}
\def\outeqn#1{\the #1}




%

\pageoffset{-0.8cm}{0.2cm}




\begintopmatter  

\vglue-2.2truecm
\centerline{\japit Accepted for publication in Monthly Notices of the R.A.S.}
\vglue 1.7truecm

\title{Power correlations in cosmology: limits on primordial non-Gaussian density fields.}

\author{A.J. Stirling$^1$ and J.A. Peacock$^2$ }

\affiliation{$^1$Institute for Astronomy, University of Edinburgh,
Blackford Hill, Edinburgh EH9 3HJ\hfill\break
$^2$Royal Observatory, \bigstrut Blackford Hill, Edinburgh EH9 3HJ\hfill\break
}

\shortauthor{A.J. Stirling, J.A. Peacock}

\shorttitle{Power Correlations in Cosmology.}


\abstract{%
 We probe the statistical nature of the primordial density 
field by measuring correlations between the power in adjacent 
Fourier modes. For certain types of non-Gaussian field, these 
$k$-space correlations would be expected to be more extended 
than for a Gaussian field, providing a useful discriminatory
test for Gaussian fields. We apply this test to the combined 
QDOT and 1.2-Jy IRAS galaxy survey and find the observed 
density field to be in good agreement with having Gaussian 
density fluctuations for modes with $k \ls 0.1 \hompc$. From 
this result we are able to set quantitative limits on a class 
of possible non-Gaussian distributions -- the product of a 
Gaussian field with an independent stochastic field. The 
maximum sensitivity is to modulations of a Gaussian field
with coherence scales of $ \sim 30 \mpcoh$ and the rms
 modulation on this scale cannot greatly exceed unity. 
We discuss the improvements to this limit
likely to be set by future surveys.

}

\maketitle  

\sec{INTRODUCTION}
A number of theories now set out to explain the initiation 
of density perturbations in the early universe. Since most 
of these ideas invoke new physics, with the prediction of 
additional fundamental fields, it is of interest to any 
cosmologist to narrow down the alternatives. One testable 
prediction that separates these theories is the statistics 
of the initial perturbations -- some predict these to be 
non-Gaussian in form (most topological defect theories, 
and some versions of inflation), and others predict that 
the density field should be Gaussian (most types of inflation).

Previous work in this area has concentrated mainly on 
three statistical approaches: (1) comparing the moments
of the density probability distribution function found
from galaxy redshift surveys with those predicted from 
a Gaussian distribution, (e.g. Saunders { et al.} 1991; Gatza\~naga 1992; 
Nusser { et al.} 1995); (2) using topology to estimate the 
genus for the observed density-field contours (for instance 
Coles et al. 1993, 
Gott et al., Moore { et al.} 1992; Park { et al.} 1992; 
Vogeley { et al.} 1994); (3) using N-body simulations to
compare the evolution and clustering properties of galaxies,
starting from different non-Gaussian and Gaussian initial
conditions (e.g. Moscardini et al. 1991, Weinberg \& Cole 1992.)

Much of the work on higher order moments has concentrated 
on measuring statistics of the density
field in the quasi-linear r\'egime, and comparing the 
observed non-Gaussian characteristics
with what would be expected for the gravitational 
evolution of initially Gaussian fluctuations. 
In the above analyses the density field has been 
found to be consistent with a distribution that was initially 
Gaussian. However, in all this work it is necessary
to smooth the galaxy distribution with some filter, 
averaging the effect of different regions of space. 
The central limit theorem then suggests that there is the 
danger that the appearance of Gaussian statistics will
always be produced, whatever the underlying distribution.

An alternative approach adopted by Feldman, Kaiser, and 
Peacock (1994; FKP) is to look at the density field in 
Fourier space, where modes are separated out on the basis 
of scale, making it easy to probe the linear r\'egime.
 FKP proposed two possible tests for a Gaussian density field. 
Both make use of the property that at high resolution in 
$k$ space, independent Fourier modes have power values 
that fluctuate about the mean power for that
particular scale. The first test was to look at the one-point
 distribution of these power fluctuations, which is 
predicted to be exponentially distributed in the Gaussian case.
 FKP found that for the QDOT 1-in-6 survey, the power 
distribution was in good agreement with the Gaussian 
prediction. However, Fan \& Bardeen (1995) have 
subsequently argued that such one-point pdf tests have 
little discriminatory power: for a large enough 
sample, most types of non-Gaussian field are also
 expected to have an exponential distribution as a 
consequence of the central limit theorem. The second 
test suggested by FKP was to look at correlations
 between the power fluctuations which, for a Gaussian field,
 will be a function of the selection function only. 
This is less susceptible to the swamping effects of the 
central limit theorem, and it is the test that
we investigate in this paper.

\sec{Power Correlations}

The general idea behind looking for correlations between 
power fluctuations is that for any infinite density field 
the Fourier modes are independent { {i.e.}}:
$$
\langle \delta{_{{\bf k}_1}} \; \delta^*_{{\bf k}_2} 
\rangle = \left(2 \pi \right){^3}
 P(  k_1)  \; 
\delta{_{\rm D}}({\bf k}_1 - {\bf k}_2),
$$ 
where $\delta{_{\rm D}}({\bf k}_1 - {\bf k}_2)$ is the Dirac
delta function and angle brackets here, and 
throughout the rest of this paper denote the
ensemble average.
For a finite sample of the density field, however, this mode 
independence is lost. This is because the finite region 
is effectively a product of the true infinite field, 
$\delta{^{\rm infinite}} \;$, and a mask, $m$. In
 Fourier space the observed density modes 
$\delta_{\bf k}^{\rm observed}$
 are then a convolution of the `true' density modes,
 with the Fourier transform of the mask.
$$
\delta_{\bf k}^{\rm observed} = \delta_{\bf k}
^{\rm infinite} \star m{_k}
$$
where $\delta_{\bf{k}}$ is given by:
$$
\delta_{\bf{k}} = {\int \delta({\bf{r}})\; 
e^{i{\bf k \cdot r}}\; d^{3}{\bf{r}}},
$$
\noindent and $\delta({\bf{r}}) \equiv \lbrack\rho({\bf{r}})
 - {\overline \rho({\bf{r}})}\rbrack/
{\overline \rho({\bf{r}})}$ is the overdensity at position $\bf r$.
This has the effect of mixing information between 
the modes over a scale comparable to one over the $k$-space
width of the mask. Now, for a Gaussian field, the 
following relation applies (see FKP appendix B):
$$ \eqalign{
|\langle \delta{_{{\bf k}_1}}\; \delta^*_{{\bf k}_2} \rangle | {^2}
 &=  \langle  {\hat P}({\bf k}{_1})   \; {\hat P}({\bf k}{_2}) 
\;\rangle -  P({\bf k}{_1})\;  
 P({\bf k}{_2}) \cr
&= \langle \,\delta P\left({\bf k}_1\right) \; 
\delta P \left( {\bf k}_2 \right)
 \;\rangle \cr}
$$
\ineqn{\blogs}%
where $P(k)$, the ensemble-average power, comes from equation (1);
${\hat P}(\bf{k})$ is the power in
a single mode:
$$
{\hat P}(\bf{k}) = | \delta_{\bf{k}} | ^{2} ;
$$
and $ \delta P\left({\bf k}\right) = {\hat P}\left({\bf k}\right)
 -  P\left( k \right)  $. This relation is dependent
on two properties of Gaussian fields -- a normal 
distribution for the one-point pdf, and the independence
of the Fourier modes (i.e. the modes should have
uncorrelated phases). For a finite sample of a Gaussian
field, there will thus be correlations between the 
different power modes, with a two-point function
for the power fluctuations which is the square of the 
two-point function for the Fourier amplitudes. 
For non-Gaussian fields in which intrinsic phase
correlations exist, this proof does not apply, and we 
consider in Section 4 specific non-Gaussian models
in which the power correlations are broader than the 
Gaussian prediction. This does not prove that all
conceivable non-Gaussian fields could be detected
in this way, but it does show that the correlations
of the fluctuating power field give a necessary condition for
the field to be considered Gaussian.
 Figure 1 shows
how the measured field appears for the IRAS-galaxy data described 
in Section 3. 

Quantitatively, we can define a power correlation function, 
$\xi_P$ as follows:
$$
\xi_P(\Delta {\bf k}) = {{\langle \;\lbrack {\hat P}({\bf k}) -
 P(k)\rbrack 
\; \lbrack {\hat P}({\bf k}-\Delta {\bf k}) - P(k-\Delta k)\rbrack \; 
\rangle}\over{\langle \; \lbrack {\hat P}({\bf k}) - P(k)\; 
\rbrack \; ^2\rangle}} 
$$
\ineqn{\ernie}%
This is simply $ |\langle  \delta_{{\bf k}_1} 
\delta^*_{{\bf k}_2} \rangle |^2 $
 as in equation (\outeqn{\blogs}), but normalised so that 
$ \xi_P(0) = 1$. We now show how to evaluate this function
for practical datasets.

\beginfigure{1}
\epsfxsize = \hsize
\epsfbox[144 81 471 715]{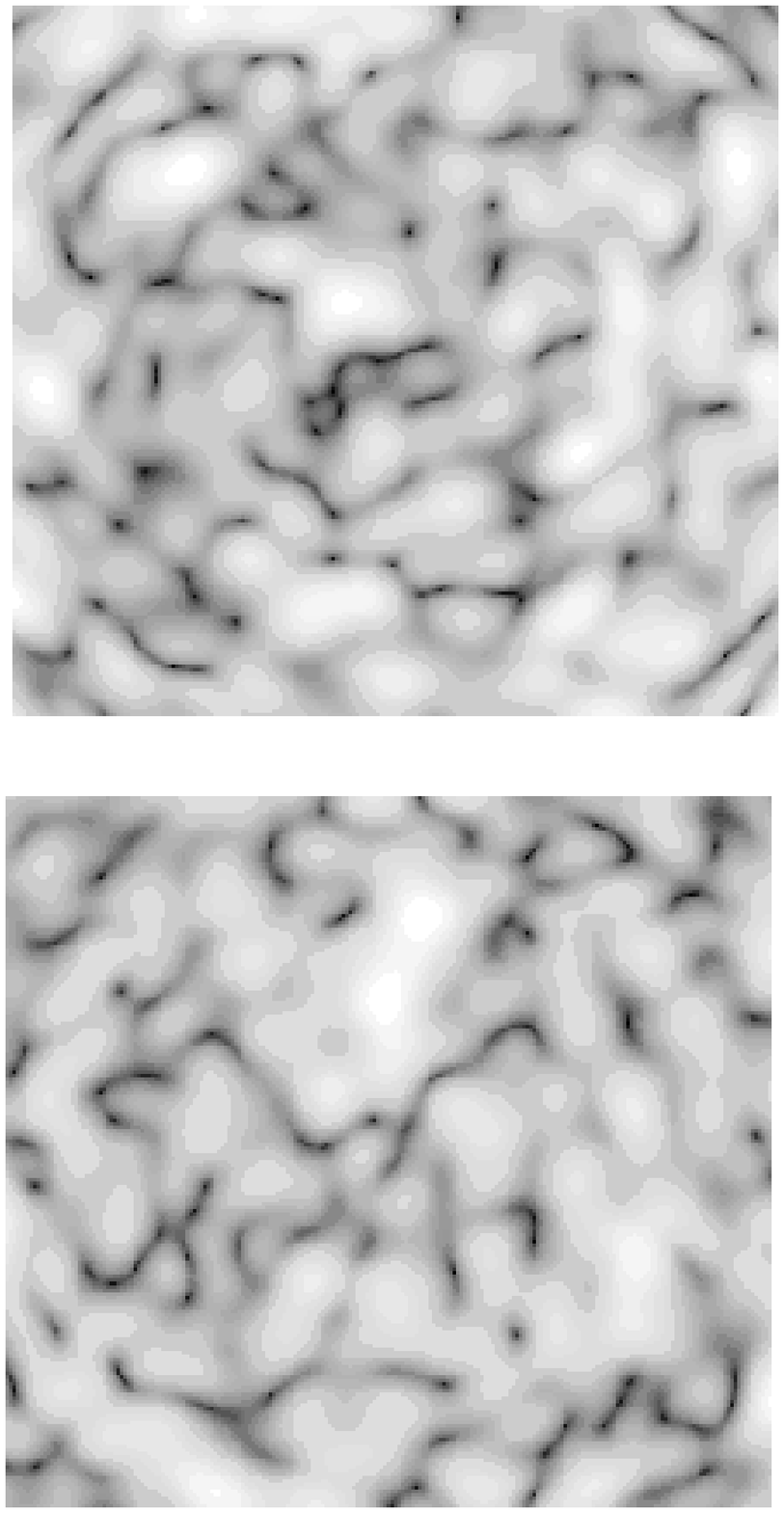}
\caption{%
{\bf Figure 1.}
Illustrating the fluctuating power 
field from the IRAS data described in Section 3. Equal area
 projections of power from 
 shells of constant $|{\bf k}|$ are shown
(Figure 1a at $k = 0.1\hompc$, and
figure 1b at  $k = 0.15\hompc$). Black corresponds to
low, and  white to high power. The panels have width $\Delta k = 0.25 
\hompc$.  While the power is not
constant for a particular value of $|{\bf k}|$,
there is a certain coherence scale over 
which there are significant correlations in power. As 
expected, this coherence scale does
not appear to change with differing values of $|{\bf k}|$, and
it forms the basis for a test for Gaussianity.
}
\endfigure
For a galaxy survey FKP defined the `weighted galaxy
 fluctuation field', $\delta({\bf r})$ as
$$
\delta({\bf r}) \equiv 
{{w({\bf r}) \lbrack n_g({\bf r}) - \alpha n_s({\bf r}) 
\rbrack }\over
{\lbrack \int d{^3}r \; {\overline n}\;{^2} ({\bf r})\; 
w{^2}({\bf r}) \;\rbrack\; {^{1/2}} }}
$$
where $n_g({\bf r}) = \Sigma_i \; \delta{_{\rm D}}
({\bf r} - {\bf r}_i)$ with ${\bf r}_i$
being the galaxy position vectors. The smooth background density 
is subtracted
via the analogous number density $n_s$,
which applies to a synthetic catalogue with number
 density $1/\alpha$ times 
that of the real catalogue. $\overline n({\bf r})$ is the
 expected mean density
of galaxies given the angular and luminosity selection
 criteria, and $w({\bf r})$ is a weighting function 
designed to minimise the variance in power by 
favouring distant galaxies while the shot noise 
is not a dominant contributor
to the power. We have used the weighting function from FKP, which is:
$$
w({\bf r}) = \lbrack 1 + {\overline n}({\bf r}) P(k) \;\rbrack {^{-1}}.  
$$

If this survey has an underlying Gaussian distribution, then FKP 
showed that the power correlation function would be of the form:


$$
\xi_P\left( \Delta {\bf k} \right) = {{{\mid  P\left( k\right) Q\left( 
\Delta {\bf k} \right) + S\left( \Delta {\bf k} \right) \mid }^2}\over {
 \lbrack P\left( k \right) + S\left(0 \right) \rbrack \; ^2}},
$$
\ineqn{\fred}%

\noindent
provided the mean power, $P$ can be taken as constant over
the width of Q. 
This expression contains two functions
of the mean density:
 
$$ 
Q\left(  {\bf k} \right) \equiv {{\int d{^3}r \; {\overline n}
\;{^2}\left( {\bf r} \right)
\; w{^2}\left( {\bf r} \right) \; e{^{i{\bf k}\cdot{\bf r}}}
 }\over {\int d{^3}r \; {\overline n}\;{^2}\left( {\bf r} \right)
\; w{^2}\left( {\bf r} \right) } }
$$
\ineqn{\bert}%
 and
$$
S\left(  {\bf k} \right) \equiv {{\left( 1 + \alpha \right) \int d{^3}r \;
 {\overline n}\left( {\bf r} \right)
\; w{^2}\left( {\bf r} \right) \; e{^{i{\bf k}\cdot{\bf r}}} }\over
 {\int d{^3}r \; {\overline n}\;{^2}\left( {\bf r} \right)
\; w{^2}\left( {\bf r} \right) } }.
$$
\ineqn{\berny}%
The idea now is to estimate $\xi_P$ via equation (\outeqn{\ernie})
and to compare this with the Gaussian prediction quoted in equation
(\outeqn{\fred}).

\sec{Application to IRAS Galaxies}
\ssec{Method}
We have found the power correlation function, $\xi_P$ for a 
combined dataset of $\sim 4500$ IRAS galaxies, consisting 
of the QDOT 1-in-6 0.6-Jy 
redshift survey 
(see Efstathiou et al., 1990; Lawrence et al. 1996) 
and the Berkeley 1.2-Jy  redshift survey
(Fisher et al. 1993, Fisher et al. 1995).
The power in each Fourier mode, ($=| \delta _{\bf k} | ^2$) 
was found following the procedure described in FKP
(see also Tadros \& Efstathiou 1995). 
 We have used a direct Fourier transform of the survey, choosing
 $\sim 6000$ random orientations in the $k$-space shell over 
which to perform the integration.
This was to sample power modes close enough together
in $k$ space to determine the shape of the correlation
function accurately.
For an estimator of $P(k)$, we have averaged 
${\hat P({\bf k})}$ over these different orientations 
in the $k$-space shell:
$$
P(k) \simeq {{1}\over{M\;N}}\sum_{i=1}^{M}\sum_{j=1}^{N} 
{\hat P}(k,\theta_i,\phi_j).
$$

\noindent $\xi_P$ was then evaluated using the assumption that the
ensemble average can be replaced by an average over 
different orientations in $k$ space. This assumption
holds for a Gaussian field, which is the 
hypothesis we will be testing against.  There is
a small amount of anisotropy introduced by the angular
part of the mask, so in comparing $\xi_P(\Delta k)$ with 
its theoretical prediction, we have also averaged $Q$, and $S$ 
over the different orientations in the $k$-space shell.
\beginfigure{2}
\epsfxsize = \hsize
\epsfysize = \hsize
\epsfbox[17 12 471 711]{fig2.ps}
\caption{%
{\bf Figure 2.} 
 The power correlation function for the QDOT + 1.2-Jy
 survey. The solid line is the theoretical prediction for this survey if the underlying 
distribution is Gaussian. Error bars are based on 
Monte Carlo simulations of the deviations
from the theoretical form for Gaussian surveys of the same
size as the QDOT + 1.2-Jy.
}
\endfigure

The normalisation of $\xi_P$ depends on the correct evaluation of 
the mean power, $P$, which can lead to problems if
 there are not enough independent patches in
$k$ space to get an accurate estimate. This 
can be overcome by evaluating $\xi_P$ on a shell of fixed $ | {\bf k} |$, 
for which the mean power is a constant. Then by requiring that 
$ \xi_P \rightarrow 0 $ for large
$ \Delta k $, we can find the actual value for $P$, and
renormalise accordingly.

 The values of $|{\bf k}|$ used were chosen experimentally 
using the criteria that the scales probed should be in the
 linear r\'egime, but not so large that shot noise dominates the signal. 
The $k$-space shell also needed to have large enough radius 
for there to be enough independent coherent patches over which 
to find an average power correlation function.

The galaxy coordinates that we use are in redshift space, 
and so peculiar velocities may 
affect the predicted shape of $\xi_P$. Empirically we find
this to be a small effect, in the sense that $\xi_P$
appears to be the same the radial and transverse directions,
and we will examine it in detail elsewhere.  

\ssec{Results}

\noindent Figure 2 is a plot of the power correlation function for the 
QDOT and $1.2$-Jy survey with 
the prediction for an underlying Gaussian distribution. The 
error bars are based on expected deviations from the curve
for a catalogue that has an underlying Gaussian distribution.
These have been determined by making real space Monte Carlo 
realisations of the galaxy catalogue, and finding the spread
in the shape of the correlation functions. The realisations
are generated by Poisson sampling from a given underlying 
probability distribution, giving all catalogues the same 
selection function, angular mask, and power spectrum as the
real catalogue, with a Gaussian distribution on large scales. 
The density fluctuations are generated as lognormal realisations 
(Coles \& Jones 1991) to give non-linear features on smaller scales.

\beginfigure{3}
\epsfxsize = \hsize
\epsfbox[11 9 493 737]{fig3.ps}
\caption{%
{\bf Figure 3.} 
Predicted shapes of the correlation function for a 
product of Gaussian field: $\delta({\bf r}) = \delta_G({\bf r})\lbrack
1 + \eta({\bf r}) \; \rbrack$ for varying forms of the power spectrum
of $\eta$, $\langle |\eta_k|^2 \rangle$.
In the upper graph, the modulating scale, $R_c$ is fixed at $50 \mpcoh$, and
the different lines represent different values of $\sigma^2$:
From the bottom up, the lines are for (1) zero amplitude of modulation (i.e. a Gaussian field); (2) $\sigma^2 =0.1$; (3) $\sigma^2 = 1.0$; and (4) (the 
uppermost)  $\sigma^2 = 10.0$
The lower graph has a fixed amplitude, $\sigma^2$ of $3.0$, with varying modulating scales. 
From the bottom up, line (1) is for the Gaussian prediction as
in the graph above; (2) $R_c = 150 \mpcoh$;
(3) the dashed line is for $R_c = 30 \mpcoh$; (4) $R_c = 100 \mpcoh$;
 and (5) (the uppermost line)
for $R_c = 50 \mpcoh$.
}
\endfigure

The figure shows that the QDOT+$1.2$ -Jy survey appears
 to lie within the error bars predicted for an initial Gaussian
 distribution. Quantitatively, the goodness of fit can be 
characterised by performing a $\chi ^2$-type test on the data: 

$$
\psi ^{2} = \Sigma_i {{(o_i - e_i)^2}\over {\sigma_i^2}},
$$
\noindent where $o_i$ is the set of observed data points, $e_i$ is the 
equivalent set expected for a Gaussian distribution, and 
$\sigma_i$ is the standard deviation for the observed data 
points, which we have taken to be the same as that found for the 
Gaussian case from the Monte Carlo simulations.
$\psi ^{2}$ does not follow the conventional $\chi^2$
distribution, since the data points themselves are correlated,
 making the actual number of degrees of freedom fewer than
 the number of data points.

The probability distribution for $\psi^2$ has been determined 
from a frequency plot of $\psi^2$ values found from the $100$ 
Monte Carlo simulated galaxy catalogues mentioned above.
We find a probability, 
$p(\psi^2 > \psi^{2 }_{\rm measured})$ of $63 \%$, (for the null
 hypothesis that the data set follows the Gaussian prediction). 
This is in good agreement with the Gaussian prediction, but how 
does it compare with what any 
non-Gaussian model might predict? 

\sec{Limits on non-Gaussian models}
Since the power correlation function for the Gaussian case is
 a function of the selection function only, one particular 
class of non-Gaussian models can be analysed with relative ease.
 This class is the product of a Gaussian
field with another stochastic field, uncorrelated with the former:
$$
\delta ({\bf r}) = \delta_G({\bf r}) \lbrack 1 + \eta({\bf r})\;
\rbrack
$$
This has been suggested as a possible alternative form for the 
density field, both on empirical grounds by Peebles (1983), and 
in some inflationary scenarios involving multiple scalar fields 
(e.g. Yi \& Vishniac 1993).
 Locally, this field looks Gaussian,
 but on larger scales it is modulated so that the amplitude of 
fluctuations varies in different parts of space giving rise to
 quiet and noisy regions. This is equivalent to a field
with zero skewness, but non-zero kurtosis. The one-point pdf 
is a simple test for this form of non-Gaussian behaviour;
however a simple transformation of the field would
restore a Gaussian looking pdf, but it would still be modulated
in a way which our method could detect.

We can predict the form of $\xi_P$ by treating the $(1+\eta)$ 
part of the field as part of the selection function, so that 
$$
{\overline n}w \rightarrow {\overline n}w (1 + \eta)
$$
in equations (\outeqn{\bert}), and (\outeqn{\berny}).
 Q then becomes:
$$
Q^\prime = Q + 2\; Q\star \eta_{\bf k} + Q\star \eta_{\bf k}
 \star \eta_{\bf k},
$$
\ineqn{\qdash}%
and S:
$$
S^\prime = S + S\star \eta_{\bf k}
$$
\ineqn{\sdash}%
We can now use equation (\outeqn {\fred}) to obtain the modified 
correlation function:
$$
\eqalign{
&\xi_P^\prime\left( \Delta {\bf k} \right) =\left \langle\,{{{\mid  
P(k) Q^\prime \left( \Delta {\bf k} \right) +
 S^\prime \left( \Delta {\bf k} \right)
 \mid }^2}\over { \lbrack \, P(k) Q^\prime \left( 0 \right) +
 S^\prime \left( 0\right)\rbrack \, ^2}}\right\rangle }
$$
\noindent
Substituting in the terms for $Q^\prime$ and $S^\prime$ from 
equations (\outeqn{\qdash}) and (\outeqn{\sdash}), and 
setting  $P Q^\prime \left( 0 \right) +
 S^\prime \left( 0\right) \equiv P^ \prime $ for ease 
of notation, this becomes:
$$
\eqalign{
&{P^\prime}\, ^2  \,  \xi_P^\prime \left( \Delta {\bf k} \right)
 =   \langle \, \mid PQ+S \mid ^2 \rangle  \cr
&\ +\langle \, \mid \left( 2PQ+S \right) \star \eta_{\bf k}+
 PQ \star \eta_{\Delta{\bf k}} \star \eta_{\Delta{\bf k}} 
\mid ^2 \rangle
  \cr
&\ + \langle \, \left( PQ+S \right) \lbrack \, \left( 2PQ+S \right) 
\star \eta_{\Delta{\bf k}}
+ PQ \star \eta_{\Delta{\bf k}} \star \eta_{\Delta{\bf k}}
\rbrack ^ \star \cr & \  + conj.  \rangle \cr}
$$
\ineqn{\slong}%
Now, by noting that odd powers in $\eta_{\bf k}$ average to zero, and 
that  $\langle \mid f \,\star \, \eta_{\bf k} \mid ^2 \rangle =\langle
\mid f \mid ^2 \rangle \star \langle \mid \eta_{\bf k}\mid ^2 \rangle$,
where $f$ is a function uncorrelated with $\eta$, 
we can rewrite equation (\outeqn{\slong}) as:
$$
\eqalign{
&{P^\prime}\, ^2  \xi_P^ \prime\left( \Delta {\bf k} \right) = 
\, \xi^G_P + 
\langle \, \mid 2{P(k)}\, Q + S \mid^2 \rangle
\star \langle \, \mid\eta_{\Delta{\bf k}} \mid ^2 \rangle \cr
&+ \langle\, \mid P(k)\,Q
\star \eta_{\Delta{\bf k}} \star \eta_{\Delta{\bf k}}
 \mid ^2 \rangle \cr
&+ \langle \, (P(k)\, Q+S
) \lbrack P(k)\, Q \star\eta_{\Delta{\bf k}} 
\star \eta_{\Delta{\bf k}}
 \rbrack ^\star 
+ conj\, \rangle \cr }
$$               
and this can be written as:
$$
\eqalign{
& {P^\prime}\, ^2 \xi_P^ \prime  \left( \Delta {\bf k} \right)\, = 
\xi^G_P\left( \Delta {\bf k} \right) \cr& \,  +
\langle \,  |2P(k)Q\left( \Delta {\bf k} \right)
 + S\left( \Delta {\bf k} \right)|^2 \, \rangle 
\star \langle |\eta_{\Delta{\bf k}}| \rangle ^2 \cr
&\, + {P(k)}^2 \langle \, |Q\left( \Delta {\bf k} \right)|\rangle \, ^2 \star 
\langle \; |\eta_{\Delta {\bf k}}|^2  \star  
|\eta_{\Delta {\bf k}}|^2\;\rangle \cr
&\, + 2\langle \,  Re \{ P(k) Q\left( \Delta {\bf k} \right) \;
\lbrack P(k) Q\left( \Delta {\bf k} \right) + 
S\left( \Delta {\bf k} \right) \rbrack ^\star \}  \rangle \,
 {\times} \cr
 & \ \ \ \ \ \ \ \ \ \ \ \ \ \ \ \ \ 
  \int \langle |\eta_{\bf k}|^2 \rangle \; d{\bf k}, \cr }
$$
\ineqn{\blah}%
\noindent where $ \xi^G _P $ is the Gaussian contribution to the 
power correlation function.  
The effect of $\eta$ is to broaden the correlation function by an
amount dependent on the power spectrum of the modulating field, thus 
acting like an unaccounted for part of the selection function.
Since this non-Gaussian field is a product
of a Gaussian with a modulating field, and the modulating part
can be treated as part of the selection function,  
we are still justified in our use of the approximation in 
equation (12).

Figure 3 shows different shapes of the power correlation function, 
for a range of power spectra for $\eta$. We have parameterised 
this power spectrum by supposing that 
the stochastic process, $\eta$, is white noise with an amplitude 
$\sigma ^2$, and a Gaussian cut-off on scales $R_c$, so that 
$$
 P_{\eta} = \langle \; |\eta_k|\; ^2 \; \rangle =
\left( 2\;\sqrt{\pi}\;R_c \right) ^3 \sigma ^2 \exp 
\lbrack{-k{^2}R_c{^2}\rbrack}
$$
and 
$$
{{1}\over{\left(2\pi\right)^3}}\int  P_{\eta}
\; d{^3}{\bf k} = \sigma ^2
$$
For the simple case of a small amplitude modulation
in the absence of shot noise, equation (\outeqn{\blah}) reduces
to:
$$
\eqalign{\langle \,|Q&^\prime (\Delta k)|^2 \, \rangle = \langle |Q + 2Q\star\eta_k|^2 \rangle \cr
&\ \ \ \ \ \ \ \ \ \ \  = \exp[-\Delta k^2 /4\alpha^2] \cr
&+ {{32 \alpha^3 R_c^3 \sigma^2}\over {(1 + 4\alpha^2 R_c^2)^{3/2}}} 
\exp[-\Delta k^2 R_c^2/(4 \alpha^2 R_c^2 +1)]}
$$ 
where $\alpha$ is the effective $k$-space Gaussian size of the survey.
This expression is illustrated in figure 3.

We are now in a position to constrain the amplitude and scale of 
modulation that could possibly be present in the observed density field.
We have the measured value of $\psi^2$
for the QDOT+1.2Jy survey from the Gaussian prediction, and
 can specify the largest allowed value of 
$\psi^2$ at a given level of confidence. For a given deviation,
 we can find a set of values of $\sigma$ and
$R_c$ for the power spectrum of $\eta$ which
 would give this level of departure from the 
Gaussian prediction. These contours are shown in figure 4. 
 Present data allow $\sigma \sim 3 $ modulation on $\sim 30 \mpcoh$ 
scales at the $95 \%$ level. 
 
\beginfigure{4}
\epsfxsize = \hsize
\epsfysize = \hsize
\epsfbox[12 4 479 725]{fig4.ps}
\caption{%
{\bf Figure 4.} 
Contour plot of $\psi^2$ deviations in $R_c-\sigma^2$ parameter space. The contours
represent different $\psi^2$ values for the deviation of $\xi_P$ from
the Gaussian theoretical prediction. The outermost contour is the measured deviation,
$\psi^2$ of $\xi_P$ for the QDOT+$1.2-{\rm Jy}$ survey from the Gaussian
form. This corresponds to a probability that $\psi^2 > \psi^2_{\rm measured}$
 of $63\%$ if the survey is from a Gaussian distribution.
The innermost contour indicates the region of 
parameter space for $\langle |\eta_k|^2\rangle$ that can be ruled out with
 $95\%$ confidence as deduced from the form of $\xi_P$ for the QDOT+$1.2$-Jy, and the corresponding error bars.
}
\endfigure

It is clear from figure 4 that the 
sensitivity to modulations is peaked around
scales of $\sim 30 \mpcoh$. From figure 3, we see
that this scale is set by the width of $\xi_P$
for the unmodulated field. Both larger and smaller 
scales give a smaller perturbation to $\xi_P$; in the 
former case because there are too few independent samples
within the sample volume. The latter case is more subtle:
although modulations characterised by a small length scale
give rise to broad power correlations, the amplitude of
the modulation is small compared with the amplitude of
the Gaussian field. So fluctuations in the 
power modes persist for small $\Delta k$ -- as with a Gaussian
field, and the effect of the modulation is only seen
at large $\Delta k$. We are least sensitive to large 
$\Delta k$ correlations because of the need to renormalise
 the correlation function due to the uncertainty
in the mean power, P (see Section 3.1).

\noindent

\sec{SUMMARY AND OUTLOOK}
This preliminary exploration suggests that the use of power
correlations can be a useful tool in allowing quantitative 
limits to be placed on certain types of non-Gaussian fields. 
Future work will include examining in more detail the effects
of redshift distortions, and extrapolating the technique for
use in larger redshift surveys. Since the error bounds scale
roughly as $\rm volume^{-1/2}$, we can expect to tighten the 
limits on $\sigma $ by about one order of magnitude with the 
next generation of redshift surveys. We should also be able
increase the sensitivity of this test to modulations on scales
approaching $600 \mpcoh$.
It will be interesting to compare the power of this 
test with alternative methods, such as the use of topology 
to detect non-Gaussian signatures.

\sec{ACKNOWLEDGEMENTS}
AJS was supported by a PPARC research studentship 
during this work. Thanks to Alan Heavens for helpful 
suggestions, Andy Lawrence for the use of the QDOT 
survey, and to Karl Fisher for the $1.2$-Jy data.

\sec{REFERENCES}

\beginrefs

\ref Coles P., Jones B., 1991 \mn, 238, 1
\ref Efstathiou G., Kaiser N., Saunders W., Lawrence A.,
 Rowan-Robinson M., Ellis R. S., Frenk C. S., 1990, \mn, 247, 10P
\ref Fan Z. H.,  \& Bardeen J. M.,  Phys. Rev. D 51, 6714
\ref Feldman H.A., Kaiser N.,  Peacock J.A., 1994, \apj, 426, 23
\ref Fisher K.B., Davis M., Strauss M. A., Yahil A., Huchra J. P., 1993, \apj, 402, 42
\ref Fisher K.B., Huchra J.P., Strauss M.A., Davis M., Yahil A., Schlegel D., 1995 \apjs, 100, 69 
\ref Gazta\~naga E., 1992, \apj, 398, L17
\ref Gott J. R., III, Weinberg D. H., Melott A. L., 1987, \apj, 319, 1
\ref Lawrence A., Rowan-Robinson M., Saunders W., Parry I.R., 
Xia X, Ellis R.S., Frenk C.S., Efstathiou G., Kaiser N., Crawford J., 1996, \mn in press
\ref Moore B., Frenk C. S., Weinberg D. H., Saunders W., Lawrence A., Ellis R. S., Kaiser N., Efstathiou G., Rowan-Robinson M., 1992, \mn, 256, 477
\ref Nusser A., Dekel A., Yahil A., 1995, \apj, 449,439
\ref Park C., Gott J. R., III, Da Costa L. N., 1992, \apj, 392L, 51
\ref Peebles P. J. E., 1983, \apj, 274, 1
\ref Saunders W., Frenk C., Rowan-Robinson M., Lawrence A., Efstathiou G., 1991, \nat, 349, 32
\ref Tadros H. \& Efstathiou G., 1995, \mn, 276, L45
\ref Yi I., \& Vishniac E., 1993, \apjs, 86, 333
\ref Vogeley M. S., Park C., Geller M. J., Huchra J. P., Gott J. R., III,
1994, \apj, 420, 525
\endrefs

\bye